# Reactive pulsed direct current magnetron sputtering deposition of semiconducting yttrium oxide thin film in ultralow oxygen atmosphere: A spectroscopic and structural investigation of growth dynamics

H. Arslan[1], I. Aulika[1], A. Sarakovskis[1], L. Bikse[1], M. Zubkins[1], A. Azarov[2], J. Gabrusenoks[1], J.Purans[1]

[1] Institute of Solid State Physics, University of Latvia, Kengaraga Street 8, LV-1063 Riga, Latvia
[2] Department of Physics, Centre for Materials Science and Nanotechnology, University of Oslo, P.O. Box 1048 Blindern, N-0316 Oslo, Norway

Corresponding author e-mail: **halil@cfi.lu.lv**

## Abstract

An experimental investigation was conducted to explore spectroscopic and structural characterization of semiconducting yttrium oxide thin film deposited at 623 K (± 5K) utilizing reactive pulsed direct current magnetron sputtering. Based on the results obtained from both x-ray diffraction and transmission electron microscope measurements, yttrium monoxide is very likely formed in the transition region between β-$Y_2O_3$ and α-$Y_2O_3$, and accompanied by the crystalline $Y_2O_3$. Resulting from either the low energy separation between 4d and 5s orbitals and/or different spin states of the corresponding orbitals' sublevels, the stability of monoxide is most presumably self-limited by the size of the crystal in thermodynamic terms. This behavior develops a distortion in the structure of the crystal compared to the metal oxide cubic structure and it also effectuates the arrangement in nanocrystalline/amorphous phase. In addition to this, spectroscopic ellipsometry denotes that the semiconducting yttrium oxide has the dominant, mostly amorphous, formation character over crystalline $Y_2O_3$. Our purpose, by means of the current findings, is to advance the understanding of formation kinetics/conditions of yttrium with an unusual valency (2+).

## Introduction

Because of their unusual valence band electronic configuration ($4f^{n-1} 5d^{0,1} 6s^2$; $n= 1\text{-}15$), rare earth elements (REE) have become, over the last decades, one of the major interesting research subjects in science and crucial components in high-tech industry/products such as catalysis field (e.g. Ce) [1], hybrid engines (e.g. Dy) [2], neutron detection (e.g. Sm, Eu) [3, 4], and fusion reactors/nuclear plants (e.g. Y) [5-7]. 3+ is the most common oxidation number exhibiting three phases as cubic (α phase), monoclinic (β phase) and hexagonal (ɣ phase), whereas the oxidation states of rare-earth metal could vary between 2+ and 4+ [8-13]. The aforementioned crystals exhibit total five multiform alterations in accordance with the ionic radius of REE and the temperature [10]. Arising from the multifaceted nature of trivalent oxide of Y structures ($Y_2O_3$) such as high melting point (~ 2700 K), high optical excitation threshold (~ 5.6 eV), high relative permittivity (~16) and mechanical and chemical stability, previous studies have primarily concentrated on the dielectric/catalytic properties [14-20]. Because the rare-earth elements that have the valence of 2+ exhibit an exclusive behavior, the acquisition of monoxide rare-earth metals has started to attract the attention of many research groups [21-

23]. Putting particular emphasis on three analyses, which are the study fields of the valence fluctuating state of pulsed laser evaporated SmO [24], the investigation of the first-principle study of superconductivity in LaO [25] and the first-principle calculations on superconductor LaO [26], may express the importance of the entire topic. Besides that, the first serious discussions and analyses of divalent yttrium (YO) in solid-phase, emerged in 2016 by Kaminaga, et al. [27]. To the best of the authors' knowledge, no report on divalent yttrium oxide deposition by reactive pulsed-DC magnetron sputtering has been found. This study was arisen from the lack of research on the oxidation dynamics of yttrium. The uniqueness of this study exists in the fact that first-time solid phase divalent yttrium (YO) was produced by reactive pulsed-DC magnetron sputtering. And for the first-time, refractive index dispersion of YO and $Y_2O_3$ mixture is obtained. One of the long-term implications of this study will impact the understanding of how oxide dispersion strengthened (ODS) nanoparticles are formed in fusion reactors [28, 29].

## Experiment

In the current research, reactive pulsed-DC magnetron sputtering and e-beam evaporator (for reference Y metal thin film deposition) attached to the multi-functional vacuum cluster tool, which was kept at a high vacuum level (~ 6.00 x $10^{-5}$ Pa) during the entire procedure, were operated in ISO 8 class clean room environment. The deposition chambers were baked out for 72 hours at 368 K (± 5K). In addition to this, for the case of high-temperature deposition, the substrates/substrate holder and close surrounding of the holder were heated up to 673 K (± 5K) for 6 hours. The substrates were cleaned in the ultrasonic bath with acetone, isopropanol, and DI water, respectively (15 min. each) and dried with $N_2$ gas. The samples were loaded and unloaded to/from the chambers using the robotic arm without disturbing the experiment environment. During the magnetron sputtering, the substrates were located 13 cm away from the target. The angle between the target and the substrate was fixed between 45-50°. The hysteresis loop of the reactive process, in the range of 0 to 6 sccm, was characterized in which the discharge was stabilized for 5 minutes in each step (Fig. 1). The plasma optical emission spectra (200-1100 nm) were measured by a plasus emicon mc spectrometer using the optical fibre probe which overlooks the discharge parallel ≈ 2 cm above the target surface. Residual and process gases ($P_{Ar}$, $P_{O_2}$, $P_{H_2}$, $P_{H_2O}$) were monitored (Table 4) by means of the process eye professional residual gas analyzer (mks instruments). The deposition of yttrium/yttrium oxide thin films, approximately 400 nm in thickness, were carried out using the metallic (% 99.99 purity) target onto Si (001) and soda-lime glass substrates. The deposition of the films was performed towards the increasing direction of oxygen flow demonstrated into the hysteresis loop (Fig. 1). The partial oxygen pressure was varied between 2.31 x $10^{-3}$ and 2.13 x $10^{-6}$ Pa (~ 5.33 x $10^{-1}$ Pa, total deposition pressure). In the case of e-beam evaporation, the substrate-target distance was fixed as 26 cm. The metal thin films' deposition (approximately 300 nm in thickness), at 298 K (± 5K) was carried out using the metallic (% 99.99, purity) yttrium pieces onto Si (001) and soda-lime glass substrates. The partial pressure of oxygen (residual gas) was detected ~ 2.00 x $10^{-7}$ Pa. The vacuum level/atmosphere is controlled by throttle valve which is located between turbomolecular pump (TMP) and process chamber and by inletting process gasses to the chambers in both cases (magnetron sputtering and e-beam evaporation). The substrates were rotated by 10 rpm in both deposition methods. No post-treatment was applied to the thin films. Growth parameters are given in detail, in Table 4.

Structural investigation of the thin films was carried out by a) X-ray diffraction: RIGAKU MiniFlex 600 benchtop X-ray diffractometer with Cu K$_\alpha$ X-ray source. The measurements were carried out between 5-60º (2θ) by 0.0050º steps with 0.0025º resolution (with the Bragg-Brentano theta-2theta diffraction geometry). 11001 data points were collected for per measurement (scan speed 1deg/min). b) Transmission Electron microscopy/SEM: A high-resolution scanning electron microscope/focused ion beam (SEM-FIB) system Helios 5 UX (Thermo Scientific) was operated at 2 kV using the Through-the-lens detector (TLD). For lamella preparation, samples were sputter-coated with a 30 nm thick gold layer followed by a 300 nm thick FIB deposited Pt layer to protect the surface from ion beam damage. The extracted lamellas were further analyzed by the transmission electron microscope (TEM, Tecnai G20, FEI) operated at 200 kV. Optical characterizations of the films were performed employing spectroscopic ellipsometer (Woollam RC2) in the spectral range from 210 to 1690 nm (or from 5.9 to 0.7 eV), and spectrometer (Agilent, Cary 7000) in the spectral range from 200 to 2500 nm. The thickness of the films was measured by profilemeter (Vecco dektak 150) and spectroscopic ellipsometer (Woollam RC2). The temperature-dependent electrical measurements between 90-350K, were carried out by Hall measurement system (Ecopia 5000). Chemical state analyses were carried out employing X-ray photoelectron spectroscopy (Escalab 250XI, ThermoFisher). The base pressure, with the charge neutralizer switched on during spectra acquisition, was better than $10^{-5}$ Pa achieved by rotary and turbomolecular pumps. The calibration and linearity of the binding energy scale was confirmed by measuring the positions of Ag $3d_{5/2}$, Au $4f_{7/2}$ and Cu $2p_{3/2}$ to be at 368.21 eV, 83.93 eV and 932.58 eV, respectively. The FWHM of Au $4f_{7/2}$ peak was better than 0.58 eV. The samples were loaded into the XPS chamber under the laboratory conditions (~300K, 1.013 x $10^5$ Pa (1atm)). The samples were sputter-cleaned, Ar$^+$ gun with the ion energy 2.0 keV was used for 30 seconds before the measurements. The size of the cleaned area was 2 x 2 mm; the incidence angle was 30º with respect to the normal of the surface. The size of the analyzed sample area was 650 x 100 μm. The excitation source was monochromated Al K$_\alpha$ radiation (hν=1486.60 eV), operating water-cooled anode at 150 W. The recorded spectra involve survey, core level, and valence band spectrum. The spectra were acquired sequentially. Complementary work function measurements were performed by UPS with monochromated He-I (see the supplementary material). The binding energy calibrations of adventitious carbon ($C_{1s}$) of the related samples were performed with respect to the vacuum level, as suggested by G. Greczynski and L. Hultman [30] (see the supplementary material) and all the spectra were shifted, accordingly. In addition, the depth profiles of H atoms were measured by secondary ion mass spectrometry (SIMS) using Cameca IMS 7f microanalyzer. The measurements were performed with the 15 keV Cs$^+$ primary beam rastered over the 200×200 μm$^2$ area and only the central part of the crater was used to collect the SIMS signal. The depth conversion of the recorded profiles was performed by measuring the sputtered crater in the films assuming a constant erosion rate. The vibrational spectroscopy analyses were performed by a vacuum Fourier transform infrared spectrometer (Bruker Vertex 80v) with 4 cm$^{-1}$ spectral resolution.

## Results and Discussion

**Structural Characterization**

This research is primarily separated into two sections. In the scope of the first step, the structural transformation of yttrium/yttrium oxide thin films deposited at 298K (± 5K), is investigated basically. The peaks detected for $S_2$, $S_3$, and $S_4$ samples that show metallic type electrical conductivity at approximately 31º (2θ) and the small peak (which is almost not visible for $S_3$) at approximately 33º (2θ), point out that metallic yttrium crystal arrangement is most probably accompanied by β-$Y_2O_3$ (Fig. 2a). Although, as the partial oxygen pressure level increases from 2.13 x $10^{-6}$ up to 4.06 x $10^{-5}$ Pa, the contribution from the oxide phase becomes dominant (while the peak located at approximately 31º (2θ) shifts towards 30º (2θ), the intensity of the peak at approximately 33º (2θ) increases); the thin films obtained, represent the same type electrical conductivity (Fig. 11). In the region between 2.11 x $10^{-4}$ and 3.97 x $10^{-4}$ Pa (the related samples are $S_9$ and $S_{10}$), β-$Y_2O_3$ is quite likely formed along with α-$Y_2O_3$. On the other hand, the single peak detected for sample $S_{11}$ could be responsible for simultaneous formation of α and β-$Y_2O_3$ or only α-$Y_2O_3$ (the slight shift of diffractogram(s) to lower 2θ values can be observed compared to the reference peak positions given). In the region~6 x $10^{-5}$ Pa ($S_7$), the film develops an amorphous structure. The deposition rates of the thin films are given in Fig. 2b. The results obtained from XRD revealed that, even at low pressure levels, yttrium tends to form oxide thanks to its high oxygen affinity. In the second step (which is the core of this article), in order to understand the effect of temperature and/or partial oxygen pressure level on the formation dynamics (phase kinetics), we deposited thin films at high temperature (~ 673 K) with similar partial oxygen pressure levels used in the amorphization and in fully oxidized regions. And we used metallic yttrium thin film as a reference (detailed experiments conditions are given in Table 4). The main peak we observed at 29.065º (2θ) in Fig. 3a, most likely corresponds to α phase Yttria (222) with the lattice parameters calculated as a = b = c = 10.6340 Å. Our results are compatible with the literature in which the lattice parameters were calculated a = b = c = 10.6431 Å [31-33]. In addition to that, the peaks at approximately 20.405º, 39.615º, 49.930º, and ≈ 60º (2θ) in Fig. 3a, denote the reflection of the x-rays from the corresponding planes in α-Yttria quite likely [33]. In the case of the metallic thin film (Fig. 3a), the major peak at 31.015º (2θ) indicates the formation of metallic yttrium crystal (possibly α-Y) [34, 35]. If we focus on sample $S_6$, the peak that appears at 36.415º (2θ) in Fig. 3a and 3b, most likely corresponds to YO (002) [21, 27, 36], which is consistent with the findings by Losego, et al. [37] in which YbO is deposited by molecular beam and Kaminaga, et al. [27] where YO is grown by a pulsed laser. In addition to that, the peak that appears at approximately 31º (2θ) (the shoulder between ~ 30.5 and ~ 31.7º (2θ)) in the same figure could be responsible for YO ((111), 31.319º (2θ) [27]) and/or β-$Y_2O_3$ (PDF Card; 00-044-0399). The results obtained from the current investigation, indicate the formation of nanocrystalline/amorphous YO accompanied by β-$Y_2O_3$ and/or α-$Y_2O_3$ (Fig. 3). The interplanar spacing (d) between the corresponding planes was calculated using the data obtained from XRD both in this research (R) and the literature (L) where d = 0.3072 (L) [33], d= 0.3064 nm (R) for α-$Y_2O_3$ (222), and 0.2485 nm (L) [27], 0.2465 (R) for YO (002). In the interest of clarity, further investigations on morphology and nanostructure were performed for semiconducting thin film by employing

HR-SEM (Fig. 4a) and TEM (Fig. 4b). Fig. 4b; r1, r2 and r3 demonstrate the different regions of TEM images of the lamella extracted from **S₆**. The separation between the regions is ~ 50 nm. r1′, r2′, and r3′ are the focused areas of the corresponding part (the yellow square) in each region. r1′ FFT, r2 FFT, r2′ FFT, and r3′ FFT are the fast Fourier transformations (FFT) of the designated regions. Fig. 4b r2′ FFT uncover the formation of amorphous phase that might contain different yttrium oxide phases in unison (some of the possible phases are YO, $Y_2O_3$). Fig. 4b r2 FFT shows the simultaneous formation of amorphous and crystalline phases. The calculated d-spacing (~ 0.3 nm) may, considering the interpretation of the spectroscopic ellipsometry and XRD data, correspond to YO. On the other hand, the calculated lattice spacing in Fig. 4b r1′ and r3′, could indicate the formation of other crystalline phases detected by XRD (β-$Y_2O_3$ (PDF Card: 00-044-0399) and α-$Y_2O_3$ (PDF Card: 01-089-5592)). In spite of that, detailed investigations employing high resolution TEM (HR-TEM), selected area electron diffraction (SAED) and synchrotron based x-ray diffraction (SXRD) are required to create more precise relation between the data obtained (will be obtained) from electron microscopy/diffraction and x-ray diffraction.

**Spectroscopic Characterization**

The main ellipsometric angles Ψ and Δ were measured at the incident angles from (55-85)° with 5° step. Refractive index $n$ and extinction coefficient $k$ dispersion curves were modelled using the Drude (DO), Gaussian (GO), Tauc-Lorentz oscillator (TLO) and Herzinger-Johs parameterized semiconductor (HJPS) oscillator functions [38]. The optical properties of Si/native $SiO_2$ substrates were obtained from SE measurements to clean substrates without the film. The surface roughness was modelled by utilizing the Bruggeman effective medium approximation (EMA) [39]. The optical gradient of $n$ and $k$ was calculated by dividing the film layer into sub-layers with smaller thicknesses and applying the EMA considering the films as a mixture of voids to vary the $n$ and $k$ values from the bottom to the top of the film [40]. Optical band gap $E_g$, for allowed direct transition, is obtained as a fitting parameter from TLO for **S₆** sample and from HJPS for **S₈**. The mean square error (MSE) between the modelled and experimental SE data was obtained in the range of 2-10. SE experimental data model-based regression analyses were performed by the Woollam software CompleteEASE®. The measured spectroscopic ellipsometry data for all samples are presented in Fig. 5. The absence of oscillation in (Ψ, Δ) for pure Y (**EB₆**) and for the sample with crystalline mixture (**S₆**) shows that these materials have high extinction (and absorption) coefficient $k$ and that the thickness of these films is above 300-400 nm. Yttrium has lower values of Ψ in respect to the sample with the crystalline mixture that signifies pure Y, which has lower refractive index n, and higher k in respect to the samples with the crystalline mixture. The numerous high oscillations in (Ψ, Δ) spectra for **S₈** sample, shows that this film is transparent for almost the entire spectral range with approximately 400 nm thickness. The decrease in oscillations above 3 eV, signifies that this material is starting to absorb the light, slightly. Fig. 6 illustrates the difference in the modelled spectra for $Y_2O_3$ samples, if the absorption is considered (or not considered) at $E$ below optical band gap $E_g$. The refractive index $n$ and extinction coefficient $k$ as a function of photon energy $E$ for Y, $Y_2O_3$, and a crystalline mixture of YO and $Y_2O_3$ materials, are shown in Fig. 7. Typical metallic characteristics in ($n$, $k$) curves can be seen for Y: $n$ and $k$ increase with the decrees of $E$ from 3 eV to 0.7 eV, and it has relatively high $k$ values (> 1) at the spectral range above 3 eV (black curves of **EB$_6$** in Fig. 7). The material that contains a mixture of semiconducting YO and dielectric phases of and Y$_2$O$_3$ (blue curves of **S$_6$** in Fig. 7) has $k$ values lower at the entire spectral range compared to $k$ for pure Y. Typical dielectric (n, k) curves can be seen for Y$_2$O$_3$ sample: ($n$, $k$) goes up with the increase of $E$ (red curves of **S$_8$** in Fig. 7). The values of the films' thickness ($d$), surface roughness ($Sr$), optical band gap ($E_g$), $n$ and $k$ at 550 nm for Y, Y$_2$O$_3$, and for a crystalline mixture of YO and Y$_2$O$_3$, are summarized in Table 1. Optical properties of Y, YO and Y$_2$O$_3$ found in the literature, are given in table 2. There are no reports in the literature about $n$ values for YO or other crystalline mixtures that contain YO. The optical $E_g$ value obtained from the TLO for crystalline mixture of YO/Y$_2$O$_3$ is (0.30 ± 0.21) eV and, it is in good agreement with Kaminaga, et al. work [27]. The findings of SE are supported by the absorbance spectra (Fig. 10b). The $n$ values for Y are higher, with respect to the values reported in the literature that could be related with deposition conditions which affect the film thickness, structure and quality [41]. $E_g$ obtained from HJPS for Y$_2$O$_3$ thin films, comply with the data in the literature [42-44]. The presence of the absorption for Y$_2$O$_3$ structure at the visible spectral range (between 2 to 4 eV in Fig.7b) is related to the variation of the ($n$, $k$) within the depth of the film due to the inhomogeneities, e.g., porosity, defects, compositional variation in the film (Fig. 8). Absorption above zero at visible range was also observed in the work of Mudavakkat, et al. [44] and Kaminaga, et al. [27] and attributed to defect states in the band gap. The decrease of $n$ toward the surface of the films is related to the increase of the film porosity. The XRD analyses of **S$_6$** evidenced the presence of crystalline mixture (Fig. 3). Thereby, the ($n$, $k$) dispersion curves for **S$_6$** should represent the effective ($n$, $k$) values as a mixture of both Y$_2$O$_3$ and YO phases, theoretically. Based on that, the interpretation of SE data as EMA mixture of Y$_2$O$_3$ and YO (Model 1) and Y$_2$O$_3$, Y and YO (Model 2) was performed to obtain (1) the volume fraction of these mixtures and (2) to obtain physical ($n$, $k$) dispersion curves for YO. Two EMA models are illustrated in Fig. 9. The introduction of pure Y in one of the EMA models was carried out to check the reliability of the simulation. Both models gave the same fit and the same MSE to the experimental data (Table 3). The volume fraction of YO obtained by SE is high with respect to the values one could expect from XRD data: the peak of YO phase has low intensity with respect to Y$_2$O$_3$ phase. High volume fraction of YO obtained by SE for the sample **S$_6$** could be explained by the dominant presence of amorphous YO phase. However, the extinction coefficient obtained for YO in both models gave comparable values to pure Y. Such observation could suggest that the film should contain a small amount of metallic Y (like in Model 2, for example), which is a contradictory information to XRD data. Therefore, the effective ($n$, $k$) curve previously obtained for **S$_6$** sample (Fig. 7), is a close representation of the physical ($n$, $k$) values for YO. The contribution of Y$_2$O$_3$ to the SE spectra of **S$_6$** is not evident: there are practically no oscillations of ($\psi$, $\Delta$) compared to Y$_2$O$_3$ sample **S$_8$**. Sample **S$_6$** has a rather higher thickness (~ 500 nm) and higher $k$ values ~0.5-1 at the entire spectral range that corresponds to high absorption values ~$10^5$ cm$^{-1}$. Thus, SE is not sensitive to the whole thickness of **S$_6$** sample, since the light is not passing through the entire depth of the film. Both high $k$ and high thickness are affecting ($\Psi$, $\Delta$) spectra which has quite low oscillations. It could be speculated that the presence of Y$_2$O$_3$, seen in XRD spectra, is pronounced more for the first 100-200 nm of the film, which is "invisible" for SE due to high $k$ of $S_6$ sample near the surface (last 100-200 nm). On the other hand, the contribution from $Y_2O_3$ cannot be seen in the $k$ curves obtained (Fig. 7b) and also in the corresponding absorption $\alpha$ curves (Fig. 10), too. This could also be related to the fact that the $E_g$ values $Y_2O_3$ are approximately 6 eV and thus $k$ and $\alpha$ are not still increasing rapidly. In the work of Kaminaga, et al. $\alpha$ has already reached $\sim 13 \times 10^5$ cm$^{-1}$ at approximately 6 eV, while in our case $\alpha$ reaches $\sim 10^5$ cm$^{-1}$ at same proximity, which is one order less. Thus, $\alpha$ curves of the sample with crystalline mixture by Kaminaga, et al. are also affected by $Y_2O_3$ contribution, probably due to lower $E_g$ values of $Y_2O_3$, and high volume fraction of $Y_2O_3$ in the sample. It should be noticed that; $\alpha$ values for $S_6$ are higher with respect to Kaminaga, et al. (Fig. 10) due to a major contribution from the YO phase that supports the conclusions already drawn from EMA simulations, considering XRD data and SE analyses carried out aforehand. It could be concluded that; effective ($n$, $k$) curves obtained for $S_6$, are significantly dominated by the amorphous semiconducting YO, and are close representations of the physical ($n$, $k$) values for YO. Nevertheless, additional studies on thinner samples ($\sim$200 nm) fabricated under the same conditions as $S_6$, will help to have a deeper understanding about the YO and $Y_2O_3$ phase formation dynamics, volume fraction and physical ($n$, $k$) dispersion curve characteristics for both crystalline and amorphous YO.

The main purpose of employing XPS in this research is the chemical state identification of yttrium (Y3d) in $EB_6$, $S_6$, and $S_8$ samples. Fig.12a represents Y3d spectrum that belongs to $EB_6$; the peak detected at approximately 155.90 eV, points out the formation of $Y^0$ (Y3d$_{5/2}$). The result obtained is consistent with the findings of the previous study by Mongstad,T. et al. [45]. Additionally, Fig. 12a′ is the O1s spectrum of the coinciding Y3d spectrum. As it can be clearly understood, no peak has been detected. Fig. 12b shows Y3d spectrum of the sample named as $S_6$. The peak observed at 156.49 eV signifies the formation of $Y^{2+}$ (Y3d$_{5/2}$). The results obtained are compatible with the literature findings 156.40 eV (YO) [27] and 156.50 eV (YH$_{2.1}$) [46]. Further, at the same spectrum, the peak observed at 156.81 eV refers to the formation of $Y^{3+}$(Y3d$_{5/2}$) [47, 48]. Fig. 12c shows Y3d spectrum of sample $S_8$ (based on the results obtained in the bulk of the article of interest, $Y_2O_3$ must be the utmost predominating phase in sample $S_8$ that requires Y 3+ formation), although the main peak detected at around 156.88 eV refers to $Y^{3+}$ (Y3d$_{5/2}$) which is compatible with the results obtained by Nefedov, V., et al. [48] the reason behind the small shoulder which exists on the spectrum (at approximately 155.30 eV) is not clear to the best of authors' knowledge and the current literature (one of the reasons can be the formation of different chemical state of yttrium that results from the interaction between Ar ion(s) and sample). Considering the O1s spectrum of samples $S_6$ (Fig. 12b′) and $S_8$ (Fig. 12c′); Albeit, the peak detected (both Fig. 12b′. and Fig. 12c′.) at approximately 529.20 eV, corresponds to metal-oxygen (In this study Y-O)) [48], the philosophy behind the exchange of the intensities and the positions (528.40 eV- Fig. 12b′.) and (527.50 eV- Fig. 12c′) is not clear (relying on the NIST data base). Moreover, additional SIMS measurements were performed to understand H content in the samples $S_6$ and $S_8$ (Fig. 6., supplementary material). The concentration of oxygen was measured $\sim 10^2$ higher than hydrogen concentration in both samples. We conclude, the contribution/contamination derived from H must be small enough. The results obtained from both XRD and SE, support the conclusions that come up from XPS measurements.

Vibrational spectrum investigation is carried out in Far-Ir (50-240 cm$^{-1}$) and Mid-Ir (240-700 cm$^{-1}$) region (Fig. 13), separately. In Mid-Ir region, we observed 6 peaks at ~ 240, 300, 335, 370, 460, and 555 cm$^{-1}$ based on the normal vibration mode of cubic sesquioxide yttrium lattice. The results we obtained, are consistent with the conclusions of previous studies [49-52]. We observe the peaks in both cases (sample $S_8$ (used as reference) and sample $S_6$). The intensity of the detected peaks demonstrates attenuation for sample $S_6$ compared to sample $S_8$. We interpret that the decrease in the intensity of the peaks is most likely due to the formation of other crystalline phases (in the case study both semiconducting yttrium oxide with 2+ oxidation state and/or β-$Y_2O_3$), and of amorphization. For the reason that the electromagnetic wave couples with the vibrations of phonons in semiconductors in Far-IR [53], we focus on $S_6$ thin film in this region. But, it might also lead a challenge. To cope with this problem, we address two well-known phenomena as the strain (stress) accumulated in the thin film during the growth [54, 55] and/or the formation of nanocrystalline/amorphous phase we detected in our thin film. We start with strain (stress) accumulation on the thin film. Born effective charge is influenced by strain (stress). This differentiation in the charge results in the electrical permittivity of the material of interest. Consequently, the coupling frequency of the electromagnetic radiation with both phonons and other elements, diversifies. This relation is noticeable for BaO, MgO, SrO, CaO [56]. Considering this phenomenon, the peaks in the region 90-170 cm$^{-1}$ (70-200 μm) may presumably signify the formation of YO. The observation of EuO phonons in a similar region was obtained by Goian, V., et al. [57]. The coupling of electromagnetic wave with the vibrational frequency of the absorbers is common for most of the cubic rare-earth oxides between 90–170 cm$^{-1}$ [58]. Now we can discuss the nanocrystalline/amorphous formation. Normal vibrational modes are known as phonons for crystalline structures. On the other hand, if we deal with the amorphous materials (or nanocrystals), understanding the dynamics of phonons will be formidable [59-62]. One of the most noticeable differentiation in the spectrum is the broadening of peaks [59] (Fig. 13b and a). This might generate the observation obtained by Goian, V., et al., [57] more clear and those findings could indirectly point out the monoxide formation in our thin film. In IR spectrum, the coupling of electromagnetic wave with the traverse optical phonons (TO) and longitudinal optical phonons (LO) appears in pairs for the crystalline materials. Whereas in the amorphous materials, traverse optical phonons are more noticeable than the longitudinal optical phonons [60, 63, 64]. These findings may explain the reason behind the observation of low intensity peaks in the region 90-170 cm$^{-1}$ (Fig. 13a), and 70-200 μm (Fig. 13b, the real space representation of the spectrum shown in Fig. 13a, in which, the relevant region becomes more noticeable). Our interpretations acquired from IR spectrum analyses show compliance with the results obtained from both structural and spectroscopic characterizations performed aforehand. Notwithstanding, further research must be carried out in order to construct an understanding of the crystalline size dependency of the spectrum, and of the effect of accumulated stress (strain) in the structure on the spectrum.

**Electrical Characterization**

In order to understand the transition dynamics of electrical conductivity from metallic to insulating demeanor, temperature-dependent electrical conductivity measurements were performed. The partial pressure level of oxygen was varied between $2.00 \times 10^{-7}$ and $2.31 \times 10^{-3}$ Pa (Fig. 11, and Table 4). The thin films obtained with $4.06 \times 10^{-5}$ Pa and lower partial oxygen pressure values represent the metallic conductivity. The conductivity of the films increases with the decrease of the oxide phase/oxygen concentration in the host crystal structure (possibly α-Y). This relation becomes more noticeable if Fig. 2, 3 and 11 are considered/compared simultaneously. On the other hand, the pressure levels higher than ~ $8.00 \times 10^{-5}$ Pa are suitable for fully oxidize yttrium. The most striking result to emerge from the data is that; ~$7.55 \times 10^{-5}$ Pa is found as the convenient partial oxygen pressure level for the formation of semiconducting yttrium oxide at high temperature (623 K). The current findings also indicate that the range of the partial pressure level of oxygen for the formation of trivalent yttrium ($Y_2O_3$) and divalent yttrium (YO), is very sensitive. The aforementioned finding is consistent with the literature, which indicates that, YO shows semiconducting behavior as EuO and YbO [8]. Furthermore, Kaminaga, et al. observed the semiconducting behavior for yttrium monoxide (YO) [27].

**Conclusion**

This paper primarily investigates the relation between the oxidation dynamics and structural formation (crystalline or amorphous) of yttrium oxide thin film. The results obtained, confirm the previous findings and contribute to the additional evidence which suggests that; under specific conditions, yttrium can take 2+ oxidation state in solid phase. One of the most significant findings gained from this study is the deposition of yttrium monoxide (YO) acquired from a metal target, which uses the reactive pulsed-DC magnetron sputtering for the first time. The second major finding is that the semiconducting yttrium oxide is presumably formed in the transition region between α phase and β phase yttria at high temperature, and accompanied by the crystalline $Y_2O_3$. The third finding is that, n dispersion curves are evaluated for the mixture of YO and $Y_2O_3$ for the first time. Moreover, it was demonstrated that the major contribution in evaluated (n, k) curves of YO and $Y_2O_3$ mixture, is provided by the presence of amorphous YO. Fourthly, in the case of the thin films that show metallic conductivity, the oxide phase (quite likely β-$Y_2O_3$) is formed simultaneously with metallic Y crystal, at room temperature. The current results are added to a growing body of literature on the solid-state divalent yttrium arrangement. This research has thrown up many questions in need of further investigation. Further research needs to examine the links between the oxide phase (α and/or β-$Y_2O_3$) and Y 2+ formation more closely using techniques such as XAS and theoretical modelling, and to establish whether/if obtaining thermodynamically stable single phase semiconducting yttrium oxide (YO) is possible.

# Acknowledgment

- This study was financially supported by ERDF project No. 1.1.1.1/21/A/050 "Large area deposition technologies of multifunctional antibacterial and antiviral nano-coatings".
- Institute of Solid State Physics, University of Latvia as the Centre of Excellence has received funding from the European Union's Horizon 2020 Framework Programme H2020-WIDESPREAD-01-2016- 2017-TeamingPhase2 under grant agreement No. 739508, project CAMART2.
- The Research Council of Norway is acknowledged for the support to the Norwegian Micro- and Nano-Fabrication Facility, NorFab, project No. 295864.

**Figures**

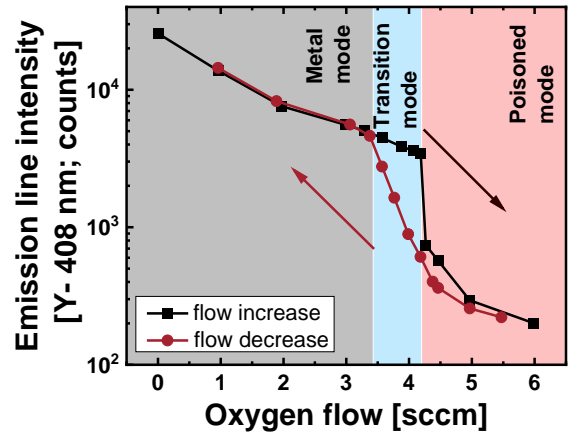

**Fig. 1.** The hysteresis loop of the reactive deposition in the range between 0 and 6 sccm.

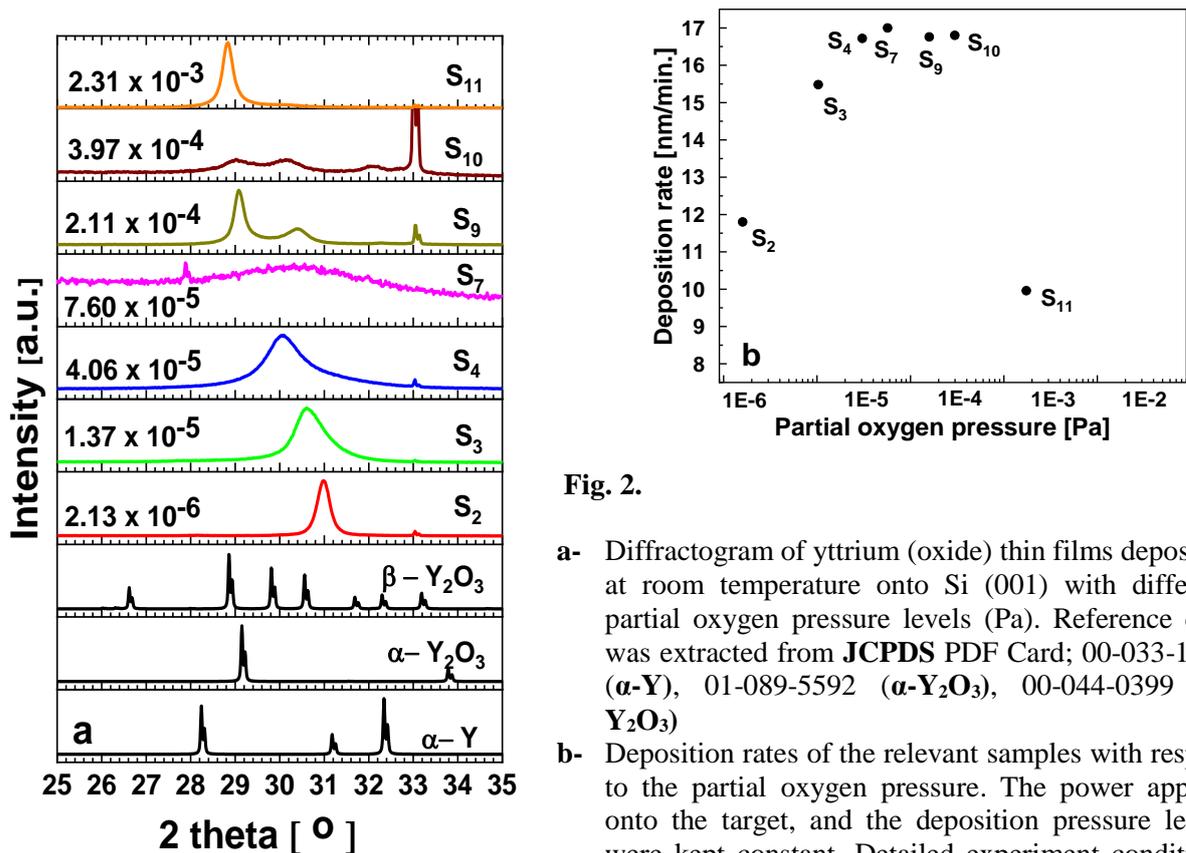

Fig. 2.

a- Diffractogram of yttrium (oxide) thin films deposited at room temperature onto Si (001) with different partial oxygen pressure levels (Pa). Reference data was extracted from **JCPDS** PDF Card; 00-033-1458 (**α-Y**), 01-089-5592 (**α-Y$_2$O$_3$**), 00-044-0399 (**β-Y$_2$O$_3$**)

b- Deposition rates of the relevant samples with respect to the partial oxygen pressure. The power applied onto the target, and the deposition pressure levels were kept constant. Detailed experiment conditions are given into Table 4.

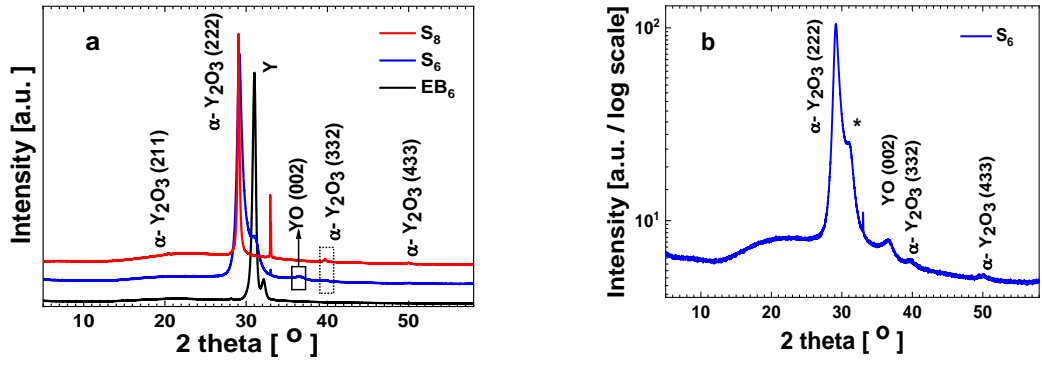

**Fig. 3.** Diffractogram of yttrium oxide that shows semiconducting behavior ($S_6$) and reference thin films: **a-** Blue line: Semiconducting thin film ($S_6$), Red line **α-Y$_2$O$_3$** ($S_8$),: Black line: metallic Y ($EB_6$); **b-** The thin film shows semiconducting behavior in logarithmic scale. The peak at approximately 31 º (2θ) may result from **YO** (111) [27] and/or **β- Y$_2$O$_3$** (PDF Card; 00-044-0399).

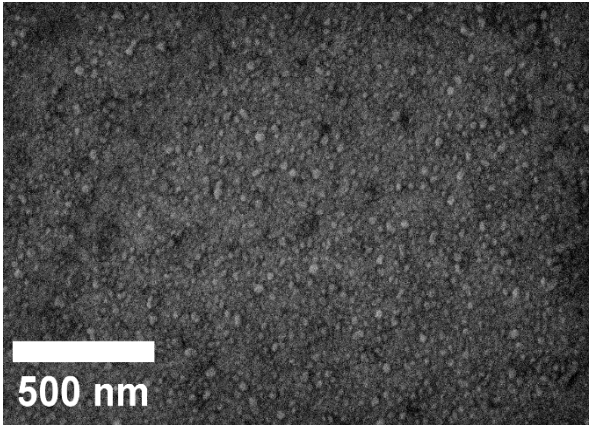

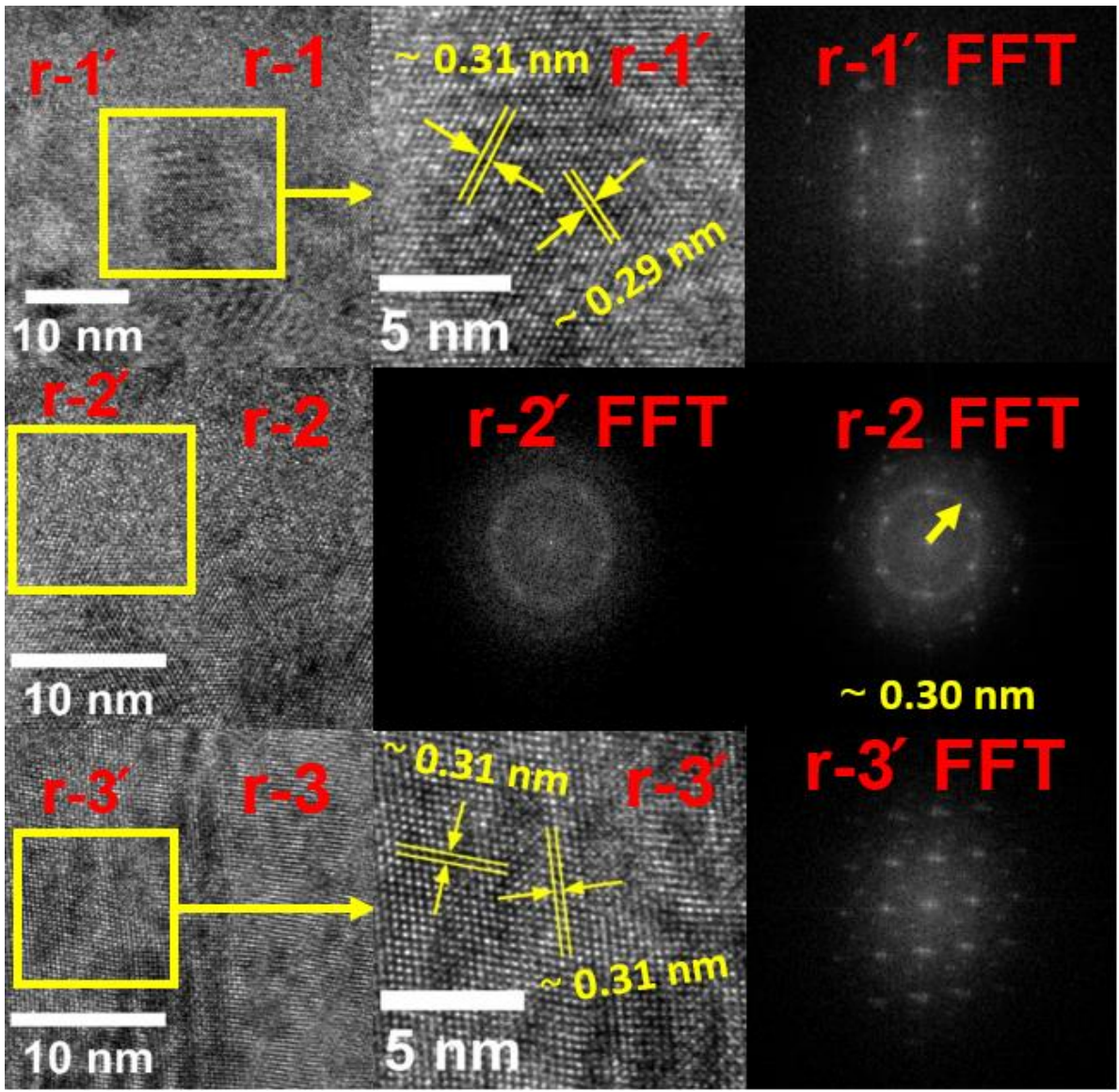

**Fig. 4.**

a- HR-SEM image of $S_6$

b- TEM images of $S_6$;

**Left column:** r1, r2, and r3 are the different regions of the extracted lamella (separation ~ 50 nm).
**Middle column:** r1′, and r3′ are the focused areas of the indicated regions in the left column and r-2′ FFT is the fast Fourier transformation (FFT) of region r-2′.
**Right column:** r-1′ FFT, r-2 FFT and r-3′ FFT are the fast Fourier transformations of the corresponding regions given in the left column.

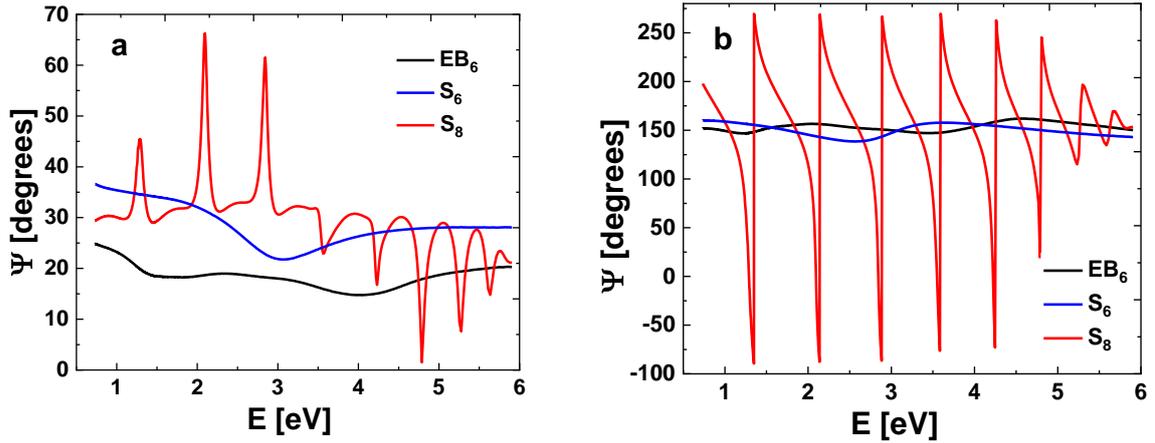

**Fig. 5.** Main ellipsometric angles Ψ (a) and Δ (b) as a function of photon energy $E$ for three different samples.

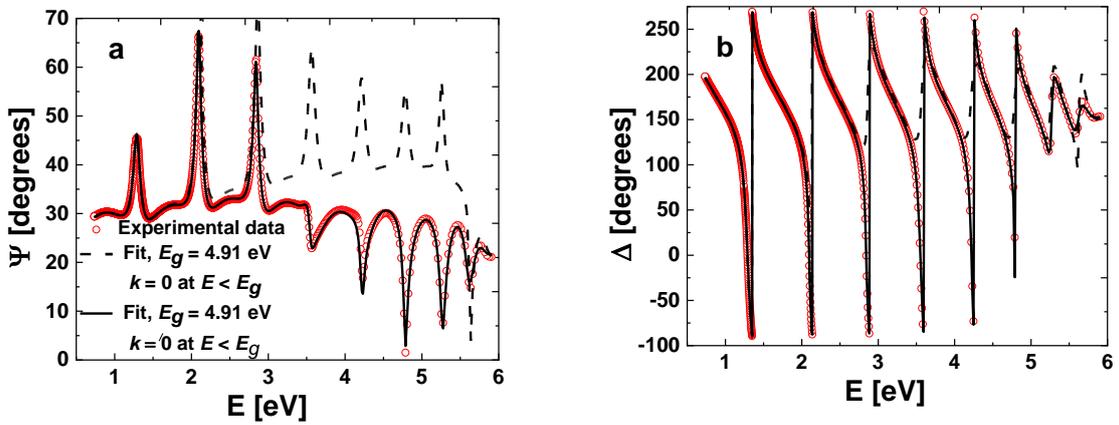

**Fig. 6.** Experimental (symbols) and modelled (lines) main ellipsometric angles Ψ (a) and Δ (b) as a function of photon energy $E$ for three different samples. Two models are given to show the difference in spectra if absorption at $E$ bellow $E_g$ is taken (or not taken) into account.

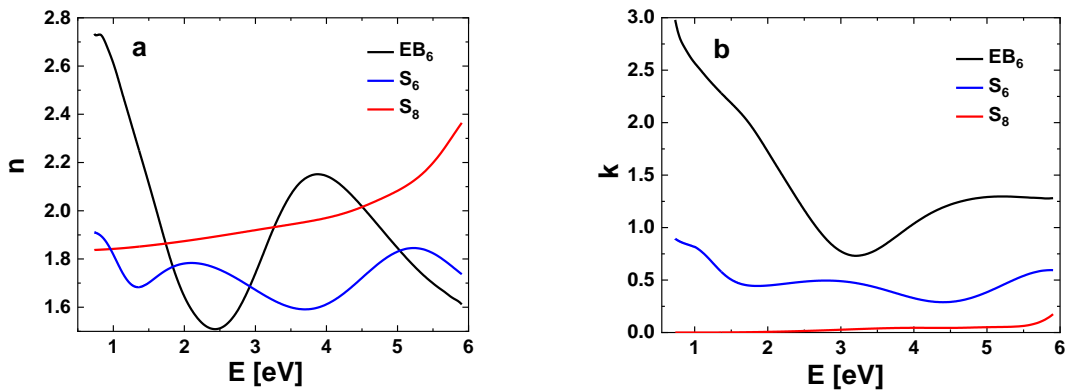

**Fig. 7.** Refractive index $n$ and extinction coefficient $k$ as a function of photon energy $E$ for three different samples: **Y** (**$EB_6$**), sample with a crystalline mixture of **YO** and **$Y_2O_3$** (**$S_6$**), and **$Y_2O_3$** (**$S_8$**).

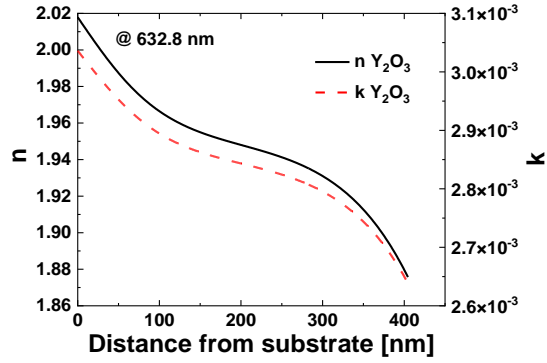

**Fig. 8.** Nonlinear depth profile for $Y_2O_3$ sample ($S_8$) at 632.8 nm (He-Ne laser) wavelength. The film layer was divided into 15 sub-layers (optimal value found for the lowest MSE).

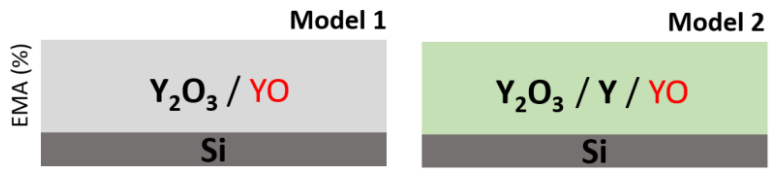

**Fig. 9.** Two different Bruggeman EMA models applied to fit SE data of $S_6$ sample to find physical ($n$, $k$) dispersion curves of **YO** and volume fraction of the mixtures. Model 1: film is a mixture of **$Y_2O_3$** and **YO**. Model 2: film is the mixture of **$Y_2O_3$**, **Y** and **YO**. In the models the obtained ($n$, $k$) curves for Y and $Y_2O_3$ where used. Resistivity values and optical band gap for **YO** was fixed to 0.045 Ω·cm and 0.1 eV, respectively, to lower the number of fitting parameters.

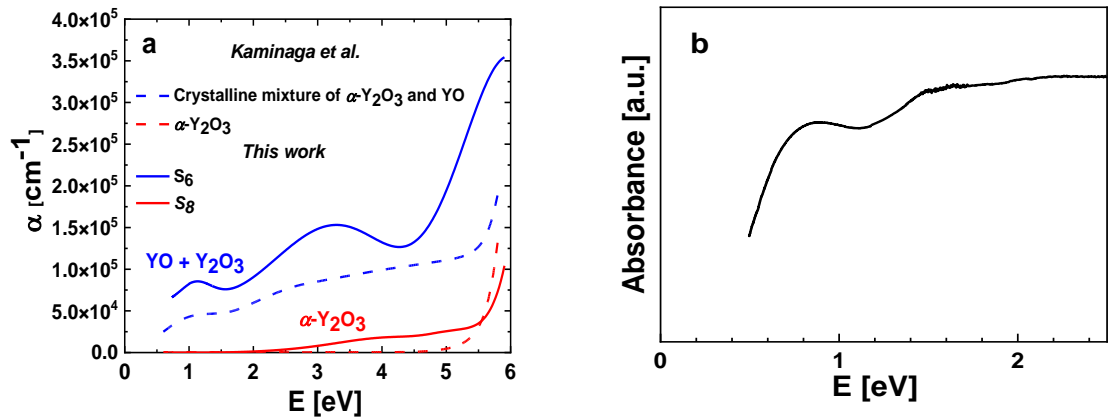

**Fig. 10.** a) Absorption curves for $S_6$ and $S_8$ samples calculated from extinction coefficient values (Fig. 7b) and compared with the results of Kaminaga, et al.[27], b) Absorbance spectra of sample $S_6$.

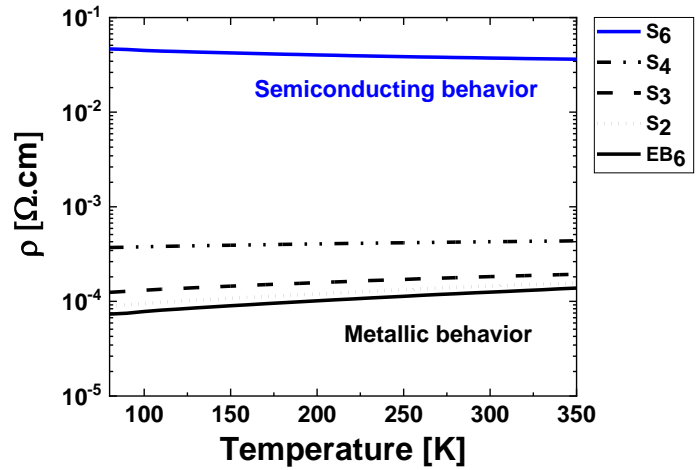

**Fig. 11.** Temperature-dependent electrical resistivity measurements of the thin films show semiconducting ($S_6$) and metallic behavior (including reference thin film ($EB_6$)) with respect to the partial oxygen pressure level. Detailed growth parameters are given in Table 4.

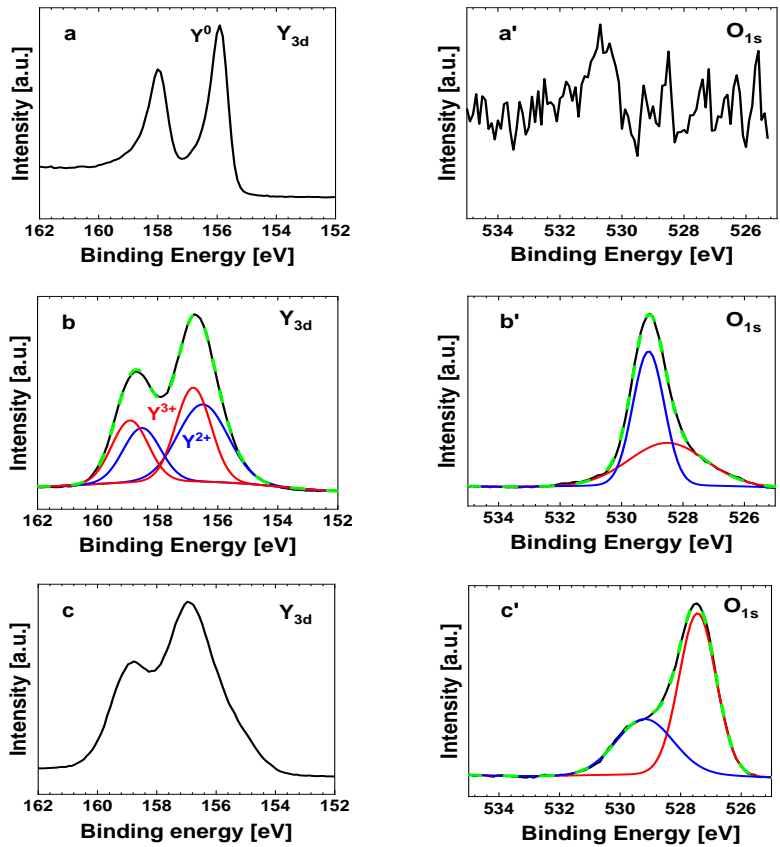

**Fig. 12.** $Y_{3d}$ (left column) and relevant $O_{1s}$ (right column) XPS measurements of the thin films that represent: a, a'- metallic ($EB_6$); b, b'- semiconducting ($S_6$), and c, c'- insulating ($S_8$) behavior. The dashed lines (green in color) in figure b, b', and c' define the fitting results of the spectrum.

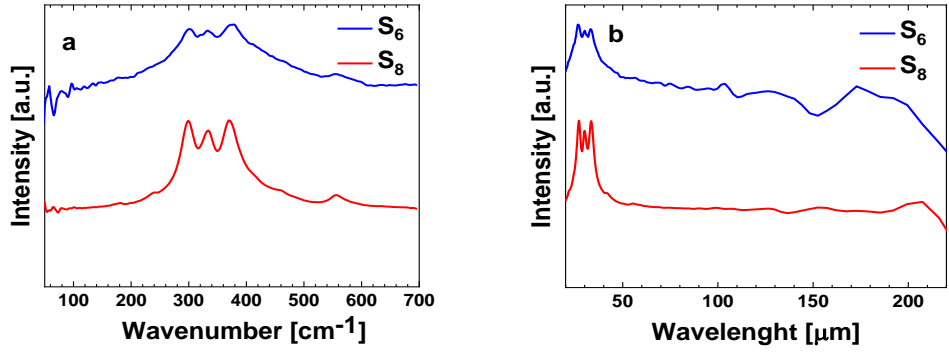

**Fig. 13.** IR Spectrum of insulating ($S_8$) and semiconducting ($S_6$) thin films, **a**-Scaled in wavenumber, **b** -Scaled in wavelength.

## Tables

**Table 1.** Obtained values of the different films on Si: $d$ - thickness of the films, $S_r$ - surface roughness, $E_g$ - optical band gap, $n$ and $k$ at 550 nm, oscillators used to model complex dielectric function of the films.

| Film | $d$, nm | $S_r$, nm | $E_g$, eV | $n$ | $k$ | Oscillators |
|---|---|---|---|---|---|---|
| Y | 200[a] | - | - | 1.533 ± 0.001 | 1.437 ± 0.001 | 1 DO, 3 GO |
| YO/$Y_2O_3$ | 500[a] | - | 0.30 ± 0.21 | 1.78 ± 0.05 | 1.47 ± 0.02 | 1 DO, 1 TLO, 3 GO |
| $Y_2O_3$ | 417.4 ± 0.5 | 12.8 ± 0.2 | 5.86 ± 0.64 | 1.89 ± 0.02 | 0.0101 ± 0.0004 | 1 HJPS, 2 GO |

[a] The film is not trispirane for the given SE spectral range: the thickness is obtained with profilometer.

**Table 2**. Refractive index at 550 nm and optical band gap for Y and $Y_2O_3$ thin films reported in literature.

| Description | | $n$ @550 nm | $E_g$, eV | Ref. |
|---|---|---|---|---|
| **Metallic yttrium Y** | | | | |
| Y thin film | | ~0.1-0.2<br>1.3 (CXRO tables) | - | [65] |
| **Yttrium monoxide YO** | | | | |
| Crystalline mixture of $Y_3O_2$ and YO fabricated by PLD on $CaF_2$ (001) substrates | | - | 0.1 | [27] |
| **Yttrium oxide $Y_2O_3$** | | | | |
| Nanocrystalline thin films grown by PVD on quartz substrates | | 1.79-1.90 (for substrate temperature during film deposition of 323-673 K) | | [66] |
| Thin films grown on to unheated [100] Si wafers with 1.2-2 μm $SiO_2$ buffer layer | e-beam deposition | 1.718 (process pressure 2.93 × $10^{-2}$ Pa with $O_2$ flow rate 70 sccm)<br>1.741 (8.00 × $10^{-3}$ Pa, 10 sccm)<br>1.917 (10.00 × $10^{-3}$ Pa, no $O_2$ flow) | - | [42] |
| | IBAD[1] | 1.911 (2.00 × $10^{-2}$ Pa, 25 sccm) | | |
| | HiTUS[2] | 1.660 (1.09 × $10^{0}$ Pa, 5 sccm)<br>1.868 (2.66 × $10^{-1}$ Pa, 9 sccm) | | |
| Bulk cubic $Y_2O_3$ | | 1.934 | | |
| Thin films on Si | | ~1.645 (973 K; ion beam, 4.00 × $10^{-3}$ Pa)<br>~1.765 (973 K; IBAD, 150 eV $O_2$)<br>~1.895 (room temperature; ion beam, none) | | [43] |
| Thin films grown by CVD on quartz and Si substrates | | 1.65 - 1.73 | 5.62 - 5.80 | [67] |
| Thin films on [100] Si wafers and quartz substrates by radio-frequency magnetron sputtering | | ~1.70 - 1.94 (substrate temperature from room temperature to 773 K) | 5.91 - 6.15 | [44] |

[1] IBAD – ion beam assisted deposition
[2] HiTUS - reactive sputtering with oxygen in a standard high target utilization sputtering system

**Table 3.** Summary of the fitting result applying two different Bruggeman EMA models to **S6** spectroscopic ellipsometry data. EMA% is the volume fraction for the corresponding material.

| Model | MSE | EMA % | | | Nr of correlating fitting parameters | $(n, k)$ curves of YO |
|---|---|---|---|---|---|---|
| | | $Y_2O_3$ | Y | YO | | |
| 1 | 8.5 | 58.9 ± 0.4 | | 41.1 ± 0.4 | 4 | $n$ lover respect to all samples; $k$ comparable to Y |
| 2 | | 48.9 ± 0.5 | 11.9 ± 0.3 | 39.5 ± 0.4 | 2 | |

**Table 4.** Deposition/Evaporation parameters of the thin films; a-magnetron sputtered thin films: **S₂, S₃,** and **S₄** are the thin films that show metallic type electrical conductivity, **S₇, S₈, S₉, S₁₀,** and **S₁₁** are the thin films represent insulating character (**S₈** is used as main reference sample), **S₆** is the thin film shows semiconducting type electrical conductivity. b- **EB₆**, e-beam evaporated thin film used as reference metal thin film. c- Partial pressure levels (approximate) of $H_2$, $H_2O$ and $O_2$ of samples **EB₆, S₆** and **S₈**.

| a | | | | | |
|---|---|---|---|---|---|
| Sample | Deposition Pressure (Pa) | $P_{Ar}$ Deposition (Pa) | $P_{O_2}$ Deposition (Pa) | Growth Temperature (K) | Power (W/cm²) |
| S₂ | ~ 5.33 x 10⁻¹ | 6.40 x 10⁻² | 2.13 x 10⁻⁶ | 298 ± 5 | 7.4 |
| S₃ | ~ 5.33 x 10⁻¹ | 4.64 x 10⁻² | 1.37 x 10⁻⁵ | 298 ± 5 | 7.4 |
| S₄ | ~ 5.33 x 10⁻¹ | 4.37 x 10⁻² | 4.06 x 10⁻⁵ | 298 ± 5 | 7.4 |
| S₆ | ~ 5.33 x 10⁻¹ | 4.92 x 10⁻² | 7.55 x 10⁻⁵ | 623 ± 5 | 7.4 |
| S₇ | ~ 5.33 x 10⁻¹ | 4.81 x 10⁻² | 7.60 x 10⁻⁵ | 298 ± 5 | 7.4 |
| S₈ | ~ 5.33 x 10⁻¹ | 4.91 x 10⁻² | 1.26 x 10⁻⁴ | 623 ± 5 | 7.4 |
| S₉ | ~ 5.33 x 10⁻¹ | 4.09 x 10⁻² | 2.11 x 10⁻⁴ | 298 ± 5 | 7.4 |
| S₁₀ | ~ 5.33 x 10⁻¹ | 3.92 x 10⁻² | 3.97 x 10⁻⁴ | 298 ± 5 | 7.4 |
| S₁₁ | ~ 5.33 x 10⁻¹ | 3.55 x 10⁻² | 2.31 x 10⁻³ | 298 ± 5 | 7.4 |
| **b** | | | | | |
| Sample | Crucible | $P_{O_2}$ Evaporation (Pa) | | Growth Temperature (K) | Current on e-Gun (mA) |
| EB₆ | W | ~ 2.00 x 10⁻⁷ | | 298 ± 5 | 43 |
| **c** | | | | | | |
| Sample | Rest/Base gases (Pa) | | | Process gases (Pa) | | |
| | $P_{H_2}$ | $P_{H_2O}$ | $P_{O_2}$ | $P_{H_2}$ | $P_{H_2O}$ | $P_{O_2}$ |
| EB₆ | ~ 10⁻⁷ | ~ 10⁻⁶ | ~ 10⁻⁷ | ~ 10⁻⁷ | ~ 10⁻⁶ | ~ 10⁻⁷ |
| S₆ | ~ 10⁻⁶ | ~ 10⁻⁵ | ~ 10⁻⁵ | ~ 10⁻⁶ | ~ 10⁻⁵ | ~ 10⁻⁵ |
| S₈ | ~ 10⁻⁶ | ~ 10⁻⁵ | ~ 10⁻⁵ | ~ 10⁻⁶ | ~ 10⁻⁴ | ~ 10⁻⁴ |